\documentclass{ws-procs9x6-cpt22}
\begin{document}

\newcommand{\refeq}[1]{(\ref{#1})}
\def\etal {{\it et al.}}
%any other macros go here 

\title{Lorentz violating inflation and the Swampland}

\author{Oem Trivedi}

\address{School of Arts and Sciences, Ahmedabad University,\\
Ahmedabad 380009, Gujarat, India}

\begin{abstract}
The swampland conjectures from String theory have had very interesting implications for cosmology and particularly for Inflation. It has been shown that the single field inflationary models in a GR based cosmology are in unavoidable tensions with these conjectures and so it becomes interesting to see whether there is a way to overcome the issues of the swampland and single field inflation in an essentially GR based cosmology. We show that this can be the case if one considers a certain type of Lorentz violating inflationary scenario. We work out the requirements for these inflationary models to be swampland consistent and then show how this scenario allows some popular inflationary potentials to be swampland consistent as well.
\end{abstract}

\bodymatter

\section{Introduction}

The concept of cosmic inflation has been extremely effective in explaining different properties of the early universe \cite{starobinskii1979spectrum,sato1981first,guth1981inflationary}. It has also been shown that a wide range of inflationary models are consistent with recent observational data. But even though inflation has been investigated in a range of exotic cosmological settings, single field inflationary models still hold a very important place in the literature. It is then very surprising that these models, which have been repeatedly shown to be observatinally consistent, are at odds with the swampland conjectures \cite{vafa2005string}. The swampland conjectures are a set of possible quantum gravity consistency criterion having their roots in string theory, which are aimed for low energy EFT's in order to check which of the vast arrays of EFT's out there today cqn eventually be consistent with quantum gravity. In particular, the dS and the distannce conjectures are in direct loggerheads with the observational requirements for inflation. \section{The swampland issues of inflation and their Lorentz violating cure} The dS conjecture is based on the proposition that no meta-stable dS spaces can be found in any consistent theory of quantum gravity \footnote{This bold claim comes from the fact that it has been very hard to find such dS spaces in the landscape of string theory } and from the point of view of scalar field cosmology, it sets a constraint on the potential (this treatment is in Planck units $m_{p}=1 $ ) \footnote{There is another "refined " form of this conjecture but we are not interested in that here} \begin{equation}
\frac{|V^{\prime}|}{V} \geq \mathcal{O} (1)
\end{equation}  
The distance conjecture on the other hand is based on the proposition that there is a limit of applicability of any low energy EFT wich is consistent with the underlying theory of quantum gravity. From the point of view of scalar field cosmology, it translates to a limit of traversable distance for the scalar field \begin{equation}
\Delta \phi \leq \mathcal{O}(1)
\end{equation}
The conjectures (1-2) set some severe implications for single field inflation. In order to have sufficient inflation, one requires that the slow roll parameters during inflation take small values (there are constant roll models as well where this is not a necessary requirement but we are not considering them here) but this requirement comes into direct issues with the dS conjecture. In particular, the $\epsilon$slow roll parameter is given by \begin{equation}
\epsilon = \frac{{V^{\prime}}^2}{2 V^{2}}
\end{equation}
During inflation one requires for $\epsilon << 1$ but it is immediately evident that this cannot be the case if one considers (1). Similarly, another observational requirement for inflation is that one requires at least 60-70 e-folds of inflation to occur for it to consistently solve the issues of big bang cosmology. It hence becomes clear that the e-folding number \begin{equation}
N = \frac{\Delta \phi}{\frac{V^{\prime}}{V}}
\end{equation}
will not even be greater than unity if one considers both (1) and (2). These are the primary swampland issues of inflation and to eradicate them, we considered a form of Lorentz violating theory given by the Lagrangian density \cite{trivedi2022lorentz} \begin{equation}
\mathcal{L} = \Bigg( \frac{R}{2 \zeta^{2}} - \frac{1}{2} \left( g{{_{\mu}}{_{\nu}}} + \xi_{1} k^{1}{{_{\mu}}{_{\nu}}}  \right) \partial^{\mu} \phi \partial^{\nu} \phi - V(\phi) \Bigg) \sqrt{-g}
\end{equation}
Using this, we arrived at a modified Klein Gordon equation \begin{equation}
\ddot{\phi } + 3H \dot{\phi} + \frac{V^{\prime}(\phi)}{1 + \kappa} = 0
\end{equation}
where $\kappa = \xi_{1} \beta_{1}$ and we call this the Lorentz violating parameter. It can then be showed that if $ \kappa >> 1 $, then inflationary models in this regime can evade both the primary swampland issues of single field inflation this essentially GR based cosmology. It can also be showed that this is true for 3 well known inflatonary models . For the Higgs potential in this regime with no free parameters,  $ \simeq  154.495$ and hence it is swampland consistent. For radion gauge inflation with a free parameter $\alpha$, it can be shown that for $ \alpha > 0.0058 $, $\kappa$ remains very much greater than unity. Finally, for a spontaneous symmetry breaking potential with 2 free parameters $ \alpha $ and $\gamma$, the requirement for $\kappa$ is fulfilled for $\alpha \leq -1 , \gamma < 0 $. The interesting thing is that these models would be inconsistent with the swampland conjectures in the usual cosmology with standard lorentz abiding General relativity and hence this work has two main takeaways.Hence the swampland criterion, and possibly by extension quantum gravity, might be pointing towards a greater role of Lorentz violations in the very early universe.
\bibliography{JSPJMJ11.bib}
\bibliographystyle{unsrt}

\end{document}